# Determination of the Boltzmann constant by laser spectroscopy as a basis for future measurements of the thermodynamic temperature

C. Lemarchand, K. Djerroud, B. Darquié, O. Lopez,
A. Amy-Klein, C. Chardonnet, Ch. J. Bordé, S. Briaudeau[♣] and C. Daussy[*]

*Laboratoire de Physique des Lasers, UMR 7538 CNRS, Université Paris 13, 99 av. J.-B. Clément, 93430 Villetaneuse, France*

[♣] *Laboratoire commun de métrologie LNE-CNAM, 62 rue du Landy, 93210 La Plaine Saint-Denis, France*

[*] *corresponding author :christophe.daussy@univ-paris13.fr, +33(0)1 49 40 33 73,+33(0)1 49 40 32 00*

## Abstract

In this paper, we present the latest results on the measurement of the Boltzmann constant $k_B$, by laser spectroscopy of ammonia at 10 μm. The Doppler absorption profile of a ro-vibrational line of an $NH_3$ gas sample at thermal and pressure equilibrium is measured as accurately as possible. The absorption cell is placed inside a large $1m^3$ thermostat filled with an ice-water mixture, which sets the temperature very close to 273.15 K. Analysing this profile, which is related to the Maxwell-Boltzmann molecular speed distribution, leads to a determination of the Boltzmann constant via a measurement of the Doppler width (proportional to $\sqrt{k_B T}$). A spectroscopic determination of the Boltzmann constant with an uncertainty as low as 37 ppm is obtained. Recent improvements with a new passive thermostat lead to a temperature accuracy, stability and homogeneity of the absorption cell better than 1 ppm over a day.

**Keywords:** fundamental constants, laser spectroscopy, absorption line shape, thermodynamic temperature

## Introduction

The Committee on Data for Science and Technology (CODATA) value for the Boltzmann constant $k_B$ essentially relies on a single experiment by Moldover *et al.* performed before 1988 and based on acoustic gas thermometry [1, 2]. The current relative uncertainty on $k_B$ is $1.7 \times 10^{-6}$ [3] (to avoid the confusion with $k$ generally admitted to be the wave vector, we denote the Boltzmann constant by $k_B$ throughout this paper). Since 1988, several projects have been developed to provide other measurements of this constant [4-7]. The renewed interest in the Boltzmann constant is related to the possible redefinition of the International System of Units (SI) expected to happen at one of the next meetings of the Conférence Générale des Poids et Mesures (CGPM) [8-16]. A new definition of the kelvin would fix the value of the Boltzmann constant to its value determined by CODATA.

At the Laboratoire de Physique des Lasers, we developed a new approach to measure the Boltzmann constant based on laser spectroscopy. The principle (proposed by Ch. J. Bordé, in 2000 [17, 18]) relies on the recording of the Doppler profile of a well-isolated absorption line of an atomic or molecular gas in thermal equilibrium in a cell. This profile reflects the Maxwell-Boltzmann distribution of the velocity distribution along the laser beam axis and leads to a determination of the Doppler broadening and thus to $k_B$. The e-fold half-width of the Doppler profile, $\Delta\omega_D$, is given by: $\Delta\omega_D / \omega_0 = \sqrt{2 k_B T / mc^2}$, where $\omega_0$ is the angular frequency of the molecular line, $c$ is the velocity of light, $T$ is the temperature of the gas and $m$ is the molecular mass.

In a pioneering experiment we demonstrated the potentiality of this new approach [19-21]. The probed line is the $\nu_2$ saQ(6,3) rovibrational line of the ammonia molecule $^{14}NH_3$ at the frequency $\nu$ = 28 953 694 MHz. The choice of such a molecule is governed by two main reasons: a strong absorption band in the 8-12 μm spectral region of the ultra-stable spectrometer that we have developed for several years and a well-isolated

Doppler line to avoid any overlap with neighbouring lines. Very encouraging preliminary results led to a value of the Boltzmann constant with an uncertainty of 190 ppm [20]. Following these firsts results, at least four other groups started developing similar experiments on $CO_2$, $H_2O$, acetylene, rubidium and cesium [22-25]. In this paper, we present recent developments of our experimental set-up and our latest determination of $k_B$.

## I. Experimental set-up

The high accuracy laser spectrometer developed for decades in our group is used here in a specific optical set-up dedicated to the measurement of the Doppler broadened absorption spectrum of $NH_3$ around 29 THz [26]. This measurement requires a fine control of: (i) the laser frequency which is tuned over a large frequency range to record the linear absoption spectrum; (ii) the laser intensity entering the absorption cell; (iii) the temperature of the gas which has to be measured during the experiment.

### *A. The spectrometer*

The spectrometer (Figure 1) is based on a $CO_2$ laser and operates in the 8-12 µm range. For this experiment, the main issues for the $CO_2$ laser are its frequency stability, frequency tunability and intensity stability. The laser frequency stabilization scheme is described in reference [27]: a sideband generated with a tunable electro-optic modulator (EOM) is stabilized on an $OsO_4$ saturated absorption line detected on the transmission of a 1.6-m long Fabry-Perot cavity. The laser spectral width measured by the beat note between two independent lasers is smaller than 10 Hz and shows an instability of 0.1 Hz ($3\times10^{-15}$) for a 100 s integration time.

Since its tunability is limited to 100 MHz, the $CO_2$ laser source is coupled to a second EOM which generates two sidebands of respective frequencies $\nu_{SB+} = \nu_L + \nu_{EOM}$ and $\nu_{SB-} = \nu_L - \nu_{EOM}$ on both sides of the fixed laser frequency $\nu_L$. The frequency $\nu_{EOM}$ is tunable from 8 to 18 GHz which is enough to scan and record the full Doppler profile. The intensity ratio between these two sidebands and the laser carrier is about $10^{-4}$. After the EOM, a polarizer attenuates the carrier by a factor 200. A Fabry-Perot cavity (FPC) with a 1 GHz free spectral range and a finesse of 150 is then used to drastically filter out the residual carrier and the unwanted sideband $\nu_{SB+}$.

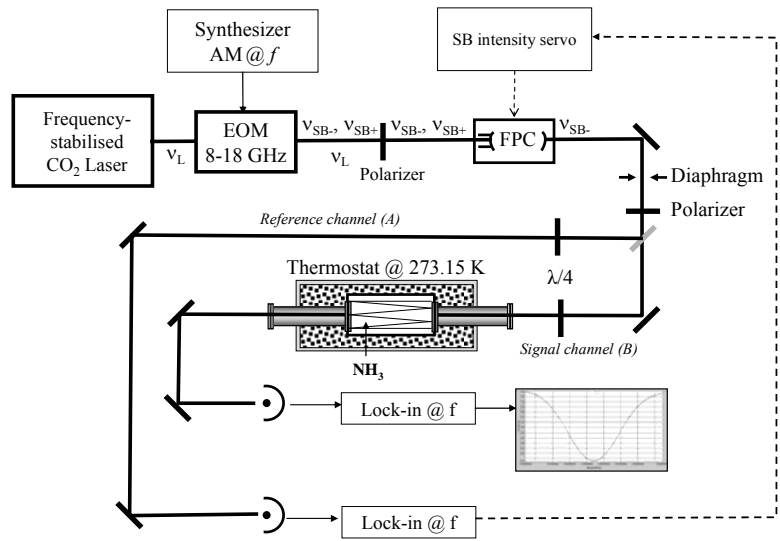

**Figure 1: Experimental set-up (AM: amplitude modulation, EOM: electro-optic modulator, FPC: Fabry Perot cavity, SB: side-band, Lock-in: lock-in amplifier)**

In order to keep the laser intensity constant at the entrance of the cell during the whole experiment, the transmitted beam is split in two parts with a 50/50 beamsplitter: one part feeds a 37-cm long ammonia absorption cell for spectroscopy (probe beam *B*) while the other is used as a reference beam (reference beam *A*). The absorption length can be adjusted from 37 cm (in a single pass configuration) to 3.5 m (in a multi-pass configuration). Both the reference beam (*A*) and the probe beam (*B*) which crosses the absorption cell are amplitude-modulated at *f* = 44 kHz for noise filtering via the 8-18 GHz EOM and signals are recovered with a demodulation at *f*. The reference signal *A* which gives the intensity of the sideband $\nu_L - \nu_{EOM}$ is compared and locked to a very stable voltage reference by acting on the length of the FPC. The probe beam (*B*) signal then gives exactly the absorption signal of the molecular gas recorded with a constant incident laser power governed by the stabilization of signal *A*. The sideband is tuned close to resonance with the desired molecular transition and scanned to record the full Doppler profile.

## *B.* *The thermostat*

The absorption cell is placed inside a large thermostat filled with an ice-water mixture, which sets the temperature very close to 273.15 K (Figure 2). The thermostat is a large stainless steel box $(1 \times 0.8 \times 0.8\ \text{m}^3)$ thermally isolated by a 10-cm thick insulating wall. The absorption cell placed at the centre of the thermostat is a stainless steel vacuum chamber closed at each end with anti-reflective coated ZnSe windows. From these windows, pumped buffer pipes extend out of the thermostat walls. They are closed on the other side with room temperature ZnSe windows. Vacuum prevents residual gas heat conduction in them and water condensation on windows.

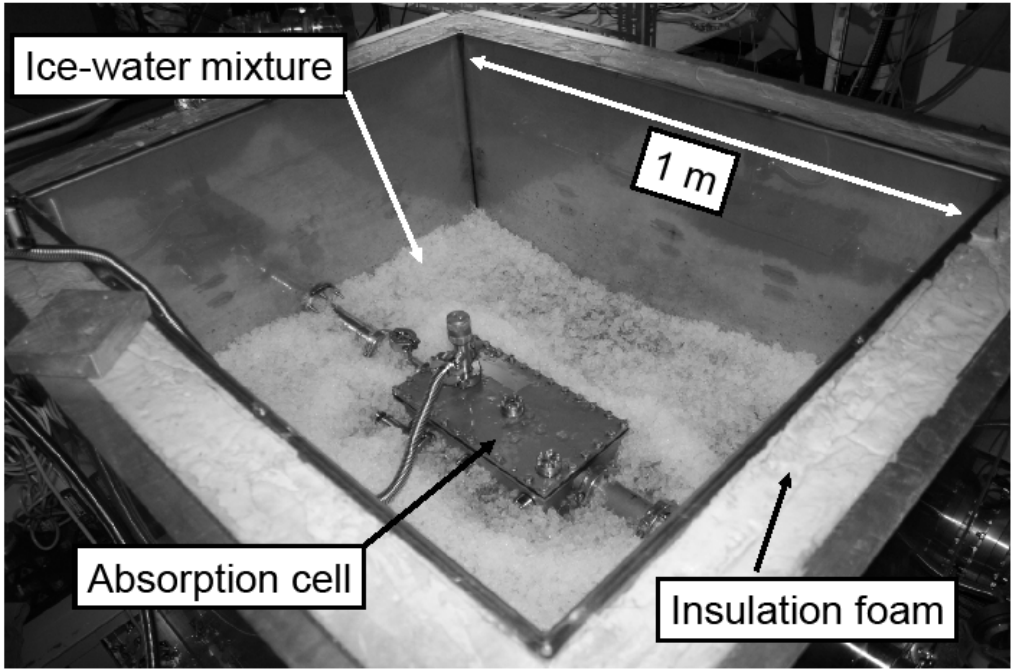

**Figure 2 : Absorption cell inside the ice-water thermostat**

Thanks to these 50-cm long pipes, the infrared laboratory Planck radiation entering the absorption cell is reduced by five orders of magnitude. The heat flux coming directly from the laboratory, $\varphi_{lab}$, can actually be estimated with [28]: $\varphi_{lab} = G\,\sigma\,T_{lab}^4$ with $\sigma$ the Stefan Boltzmann constant and $G \approx \left(\pi\phi_0^2/4d\right)^2$ (assuming $d \ll \phi_0$) the geometrical extent between the cell window (diameter $\phi_0$) and the vacuum pipe output (same diameter $\phi_0$) separated by a distance d. As $\phi_0 = 4\,\text{cm}$ and d = 50 cm, $T_{lab} = 300$ K, one finds $G = 6 \times 10^{-6}\ \text{m}^2$ and $\varphi_{lab} = 3$ mW (note that the laser intensity, below 1 µW, is more than three orders of magnitude smaller than $\varphi_{lab}$). Such a small amount of heat should be completely dissipated by heat conduction from the inner part of the cell into its outer part. It is then strongly dissipated by natural convection into the surrounding melting ice thanks to the huge enthalpy of fusion of ice. With a stainless steel conductivity of at least 10 $Wm^{-1}K^{-1}$, a cell wall surface and thickness of respectively 0.15 $m^2$ and 5 mm, we expect the temperature difference between the inner and the outer cell surfaces to remain below 0.01 mK. Moreover owing to strong heat transfer between them, the pipe and the melting ice share almost the same temperature and one can easily neglect heat conducted to the cell through the pipe from its outer part at room temperature.

The cell temperature and thermal gradients are measured with three long stems 25 Ω Standard Platinum Resistance Thermometers (SPRTs) calibrated at the triple point of water (TPW) and at the gallium melting point. The SPRTs are coupled to a resistance measuring bridge (Guildline Instruments Limited, ref 6675/A) calibrated versus a resistance standard with a very low temperature dependence. The resulting cell temperature accuracy is 1 ppm with a noise of 0.2 ppm after 40 s of integration. As shown on Figure 3, for longer integration times, temperature drifts of the cell remain below 0.2 ppm/h. The melting ice temperature homogeneity close to the cell, has been investigated with the same SPRT, simply immersed in the melting ice, and moved from one place to another around the cell, at about 1cm from the outer wall. Reproducible residual gradients have been measured. The vertical (resp. horizontal) gradient is equal to 0.05 mK/cm (resp. 0.03 mK/cm) leading to an overall temperature inhomogeneity along the cell below 5 ppm.

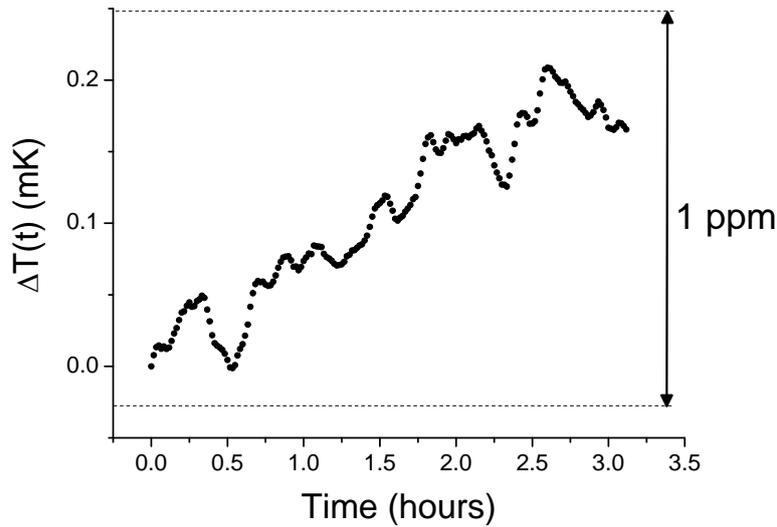

**Figure 3 : Fluctuations of the absorption cell temperature measured with an SPRT over 3 hours**

## II. The Boltzmann constant measurement

An accurate determination of the Doppler line width requires a very detailed description of the line shape. We consider the case of an optically thick medium in which the absorption is given by the Beer-Lambert law $\exp(-\alpha L)$ (actually introduced for the first time by P. Bouguer in 1729 [29]), where $L$ is the length of the absorption cell and $\alpha$ is the absorption coefficient. The absorption coefficient $\alpha$ can be described by a Voigt profile which is the convolution of a Gaussian related to the inhomogeneous Doppler broadening and a Lorentzian whose half-width, $\gamma_{hom}$, is the sum of all the homogeneous broadening contributions. Since the natural width is negligible for rovibrational levels, this homogeneous width is dominated by molecular collisions and, thus, is proportional to the pressure. In linear absorption spectroscopy and for an isotropic distribution of molecular velocities it can be shown that all transit effects are already included in the inhomogeneous Doppler broadening [30]. At high pressures, the Lamb-Dicke-Mossbauer (LDM) effect which results in a reduction of the Doppler width with pressure must be taken into account [31-34]. Various theoretical models are available in the literature, depending on the assumption made for the type of collisions between molecules [35]. Such profiles need at least one additional independent parameter to describe the line shape. Among them the Galatry profile [32] makes the assumption of so-called "soft" collisions between molecules (brownian motion) with the introduction of the self-diffusion coefficient D as a new parameter. This profile can be expressed from the correlation function of the optical dipole, denoted as $\phi_G(\tau)$ for the particular case of the Galatry profile, for which one finds:

$$\begin{cases} \alpha(\omega - \omega_0) \propto 2\,\mathrm{Re}\int_0^{+\infty} \exp(i\omega\tau)\phi_G(\tau)\,d\tau \\ \phi_G(\tau) = \exp\left[-i\omega_0\tau - \gamma_{\mathrm{hom}}\tau + \frac{1}{2}\left(\frac{\Delta\omega_D}{\beta_d}\right)^2\left\{1 - \beta_d\tau - \exp(-\beta_d\tau)\right\}\right] \end{cases} \quad (1)$$

$$\text{with } \beta_d = \frac{k_B T}{mD}$$

where $\omega$ is the laser frequency, $\omega_0$ is the frequency of the molecular line, $\beta_d$ is the so-called coefficient of dynamical friction, m the molecular mass, $\Delta\omega_D$ the Doppler width and $\gamma_{\mathrm{hom}}$ the homogeneous width. This profile evolves from a Voigt profile at low pressure to a Lorentzian shape in the high pressure limit. The self-diffusion coefficient at $P_0 = 1$ atm was chosen to be $D^0_{NH_3} = 0.15$ cm$^2$s$^{-1}$ according to measurements and calculations found in the literature [36, 37]. From $D^0_{NH_3}$, one can calculate the mean free path between collisions: $l_m = \sqrt{3m/k_B T} \times D^0_{NH_3} \times (P_0/P)$. The gas pressure in the cell ranges from 0.1 to 1.3 Pa leading to a mean free path below 1 cm.

The absorption profile of about 100 MHz wide is recorded over 250 MHz by steps of 500 kHz with a 30 ms time constant. The time needed to record a single spectrum is about 42 s. For a 100% absorption, the signal-to-noise ratio is typically $10^3$. The experimental spectra are fitted with a first-order Taylor expansion about β = 0 of the exponential of a Galatry profile (equation (1)). The Doppler widths obtained after processing 1420 spectra recorded at various pressures are shown in Figure 4, with their uncertainty (see [38] for the fitting procedure).

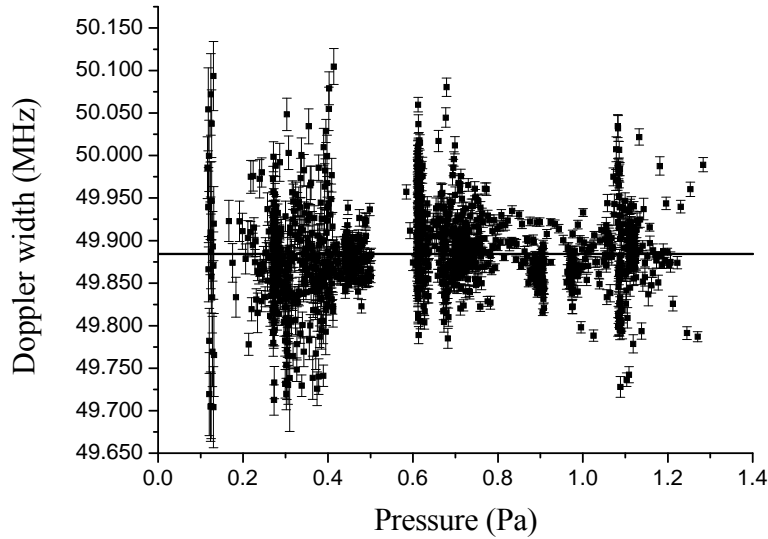

**Figure 4: The Doppler width of the sa Q(6,3) NH$_3$ absorption line versus pressure for 1420 measurements, after fitting spectra with a Galatry profile Taylor expanded to first order in β**

An average of all the data points displayed in Figure 4 led to a mean Doppler width of:

$$\Delta\nu_D = 49.884\,23\,(93)\,\mathrm{MHz}\,\left(1.9 \times 10^{-5}\right)$$

The corresponding value for the Boltzmann constant is:

$$k_B = 1.380\,716\,(51) \times 10^{-23}\,\text{JK}^{-1}\,(3.7 \times 10^{-5})$$

The 1420 spectra obtained after 16 hours of accumulation yielded a statistical uncertainty on $k_B$ reduced to 37 ppm, limited by laser amplitude noise. The value of $k_B$ was not corrected for any systematic shift and the given uncertainty of 37 ppm is only of statistical origin. Attempts to observe systematic effects due to the modulation index, the size or the shape of the laser beam, and the laser power, including non-linearity of the photodetector were unsuccessful at a 10 ppm level. No systematic effect due to the temperature control of the absorption cell is expected at a 5 ppm level (see section I.B). Systematic effects due to the "soft" collisions model chosen here to describe the LDM narrowing need to be evaluated but we are confident that the corresponding shift, assuming another collisional model, would be lower than the statistical uncertainty itself [38]. We have looked carefully at the residuals and checked that they are negligible which indicates that the experimental line shape is very close to the exponential of a Galatry profile (Figure 5).

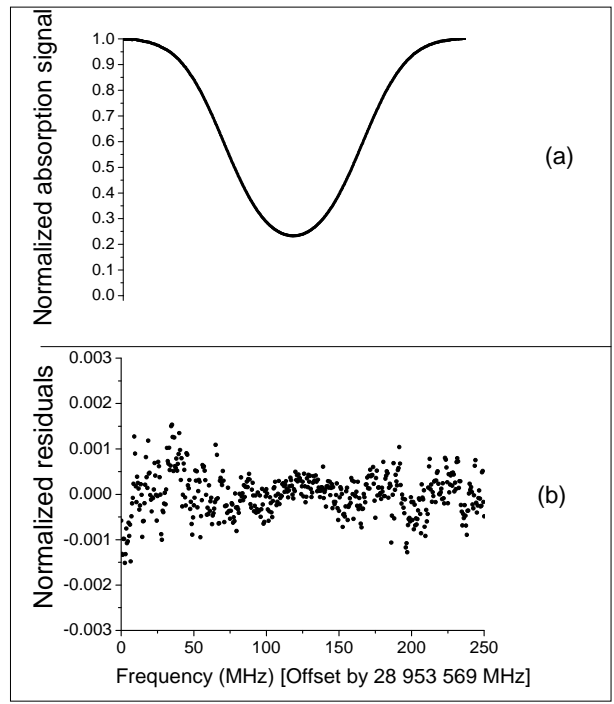

**Figure 5 : (a) Spectrum recorded at 0.9 Pa, (b) Normalized residuals for a non-linear least-squares fit of the spectrum (a) with the exponential of a Galatry profile Taylor expanded to first order in β. The frequency scale is offset by 28.953569 THz.**

## III. Development of a new thermostat with a 1 ppm temperature control

The residual temperature gradients of the ice-water mixture mainly come from the difficulty to keep a homogeneous mixture surrounding the cell. Water falls down under gravity, leaving some empty areas between bits of ice. The presence of such empty zones results in temperature inhomogeneities, bad thermal transfer between the cell and the melting ice and thus residual thermal gradients on the absorption cell. To overcome this problem, a new passive thermostat at 273.15 K is being developed.

In this new device, the absorption cell lies inside a copper thermal shield which is itself inside an enclosure immersed in the thermostat filled with an ice-water mixture (see
Figure **6**).

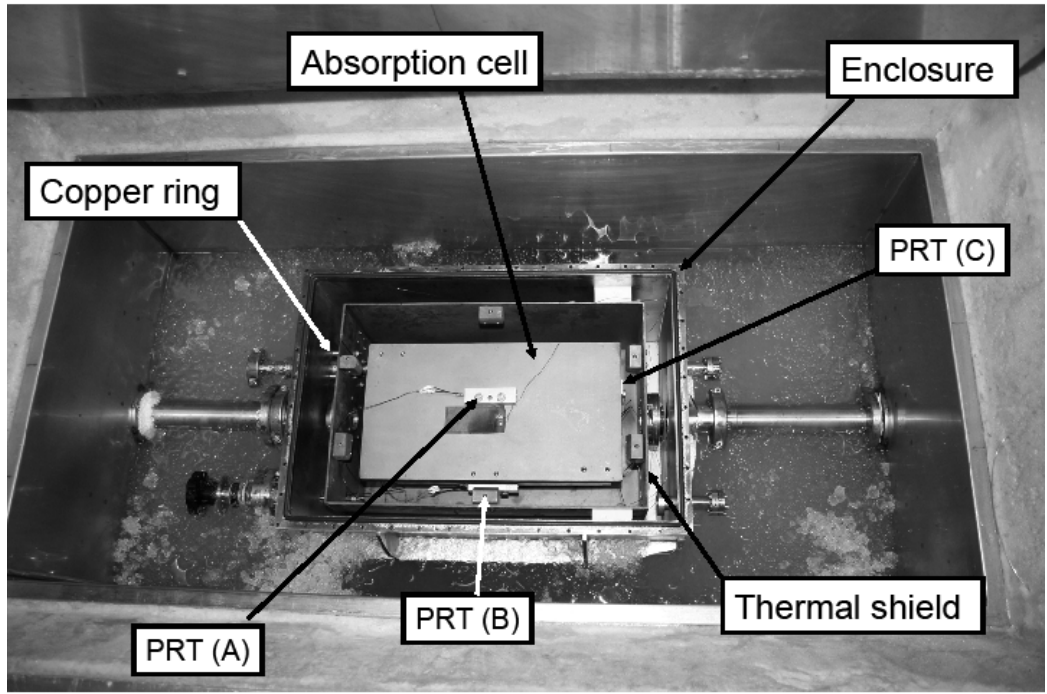

**Figure 6 : Absorption cell inside the new ice-water thermostat**
**(PRT: Platinum Resistance Thermometer)**

The enclosure containing the thermal shield as well as the pipes extending out of the thermostat, to let the laser beam go through, are evacuated. Vacuum prevents water condensation inside the enclosure and residual gas heat conduction in it. The thermal shield is solely linked to its enclosure via a tiny copper ring. The absorption cell is mechanically linked to the thermal shield with tiny Teflon holders, in order for heat transfer between them to occur mainly via Planck optical radiation. Heat transfer between these two elements is then spatially smoothed, which reduces residual temperature gradients. A manual gas valve attached to the absorption cell is used to isolate it from residual warm ammonia contained in the vacuum hose which could enter the cell. The valve handle is removable and can be managed from outside the enclosure in order to prevent residual heat conduction along the valve body. With this new set-up, thermal transfer between the absorption cell and the shield is very weak, as well as the heat transfer between the shield and the enclosure. The thermal time constant of this system is about 10 hours (at least a factor 10 larger than the previous one) which smoothes temperature drifts over the same time scales. With such a system, we expect the cell to exhibit smaller temperature gradients as well as a better temperature stability as the thermal bath temperature variations are filtered

The temperature probes used with this new experiment are small (5-cm long) Hart capsules 25 Ω Platinum Resistance Thermometers (PRTs). Three such PRTs are directly placed inside vacuum, on the absorption cell, for temperature measurements. They are associated with the resistance measuring bridge previously described. These PRTs satisfy the ITS90 criterion [39-41] although their resistance at the TPW can show a large difference with their nominal value. Due to their small size, a device has specially been designed for their calibration against the TPW. As the enclosure is immersed in the ice-water mixture, the cell shows temperature drifts smaller than 1ppm/day (see Figure 7), which is more than enough since an absorption spectrum is recorded within less than 1 min and since besides the temperature is monitored in real time.

The measurement of the absorption cell temperature using these three probes shows a 5 ppm gradient, equal to the one observed with the previous set-up. This can easily be explained. In the previous set-up, the heat transfer between the absorption cell and the melting ice was strong. In this new set-up, the 3 mW Planck heat flux entering the cell (see section I.B.) should be much more weakly dissipated towards the thermal shield through radiation. We then expect this small heat flux to create a much stronger temperature gradient on the absorption cell walls. To reduce the Planck radiation and the temperature gradient along the absorption cell, two band-pass interference filters (centred at 10 μm) were placed inside the vacuum enclosure and thermally linked to the thermal shield. In addition, a reduction of the pipe optical diameter by a factor 4 with the help of iris diaphragms (high emissivity in the cold part and low emissivity in the hot part) also reduced the geometrical extent G by a factor of 250.

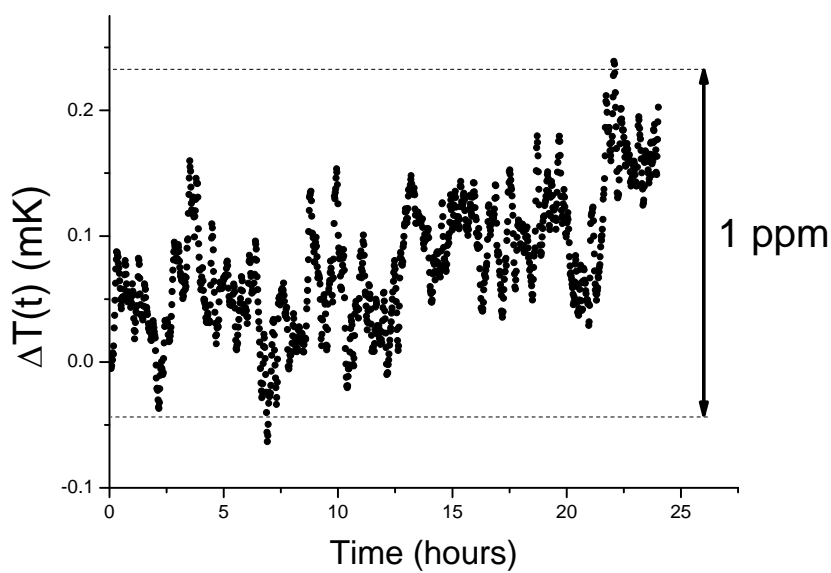

**Figure 7 : Absorption cell temperature measured with a 25 Ω PRT over 1 day**

Results reported in Figure 8 show a noise at 0.5 ppm (for ~1 min integration time). This noise is essentially limited by noise detection which can be reduced by better electrical connections between the PRTs and the resistance measuring bridge. The vertical temperature difference, between PRT (A) (at the top of the cell) and PRTs (B) and (C) (in the middle of each vertical face) is at a 0.5 ppm level. The horizontal temperature difference, between PRTs (C) and (B) is at a 0.4 ppm level. Despite the thermostat optical windows, the heat flux coming from the laboratory entering the cell has been reduced to a negligible level (compared to the 5 ppm gradient mention in the previous paragraph). To conclude, this new thermostat shows a temperature stability and homogeneity of the cell below 1 ppm (0.3 mK) over 1 day.

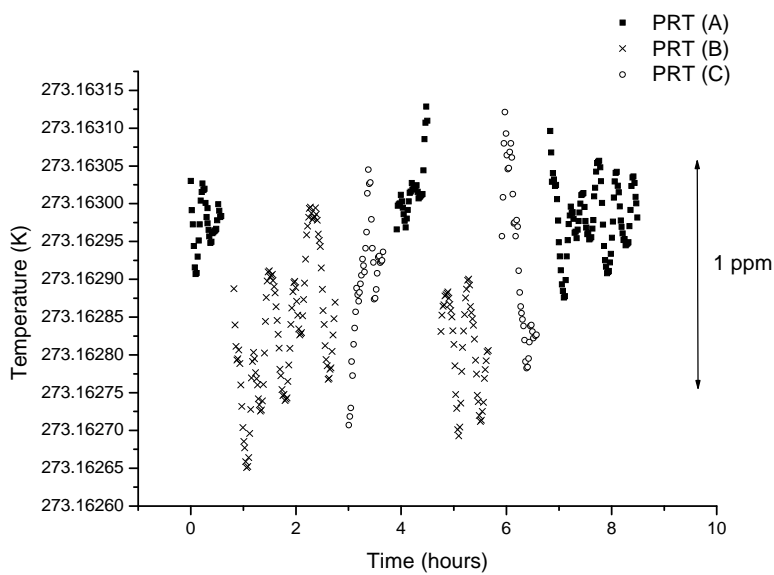

**Figure 8 : Absorption cell temperatures measured with three 25 Ω PRTs located on the absorption cell (see Figure 6)**

The main advantage of the melting ice thermostat is its simplicity and its intrinsic temperature stability. However, it is not possible to change the temperature of the bath which could be useful to perform a complete analysis of temperature dependent systematic effects. To enable a variable but stable working temperature, the melting ice will be replaced by a 1 m³ mixture of water and alcohol, maintained at a desired temperature. A cryostat actively coupled to a heat exchanger will allow a regulation of the temperature of the thermostat liquid bath from +10°C down to –10°C. In this temperature range, the relative uncertainty associated with the interpolated temperature of the PRTs used remains within 1 ppm.

## Conclusion

After a first demonstration of an optical measurement of the Boltzmann constant [19, 20], a second generation experimental set-up was designed. With this new setup, temperature control of the absorption cell below 5 ppm was demonstrated. From Doppler width measurements we deduced a value of the Boltzmann constant equal to $k_B = 1.380\,716\,(51) \times 10^{-23}\,\mathrm{JK}^{-1}\,(3.7 \times 10^{-5})$ with a statistical uncertainty of 37 ppm. This measurement is in agreement with the value recommended by CODATA, $1.380\,6504(24) \times 10^{-23}\,\mathrm{JK}^{-1}$ [3], within 47 ppm.

The main goal of this project is to perform a first measurement of the Boltzmann constant by laser spectroscopy at a few ppm level and thus to contribute to the new determination of this fundamental constant. To that purpose, a new thermostat was developed in which the absorption cell lies inside a copper thermal shield located inside a stainless steel enclosure. A cell temperature inhomogeneity and stability better than 1 ppm over 1 day was demonstrated.

Besides our main goal, a detailed study of line shape profiles at different pressures is under progress. A new 3 cm-long absorption cell has been designed to record spectra at pressures up to 200 Pa (at which the LDM effect is large) relevant to the study of molecular collisions and diffusion effects. Moreover a new multi-pass Herriott cell (37-m absorption length) will be installed on the spectrometer to record spectra at pressures lower than 0.1 Pa for which we expect the LDM contribution to the profile to be negligible for a determination of the Boltzmann constant at 1ppm level.

Finally this project could lead to the development of an absolute thermometer operating in a wide range of temperatures, from cryogenic temperatures up to several hundreds of degrees Celsius. In this range the accuracy of our experiment should exceed that of the current techniques. This is very appealing in the view of improving the temperature scale.

## Acknowledgements:

This work is funded by the Laboratoire National de Métrologie et d'Essais and by European Community (EraNet/IMERA). Authors would like to thank Y. Hermier, F. Sparasci and L. Pitre from LNE-INM/CNAM for SPRTs and PRTs calibration, discussions and advice for the thermometry part of this project.